\documentclass{aa} 
\usepackage{graphicx}
\usepackage{epsfig}
\usepackage{natbib}
\bibpunct{(}{)}{;}{a}{}{,}

\begin{document}

\title{The CORALIE survey for southern extrasolar planets\\
V. 3 new extrasolar planets\thanks{Based on data obtained with the 
{\footnotesize CORALIE} echelle spectrograph mounted on the 1.2-m 
"Leonard Euler" Swiss telescope at ESO-La Silla Observatory (Chile)}}

\author{D.~Naef\inst{1} \and M.~Mayor\inst{1} \and F.~Pepe\inst{1} 
\and D.~Queloz\inst{1} \and N.C.~Santos\inst{1} \and S.~Udry\inst{1} 
\and M.~Burnet\inst{1}}

\institute{Observatoire de Gen\`eve, 51 ch.  des Maillettes, 
CH--1290 Sauverny, Switzerland}

\offprints{Dominique Naef,
\email{dominique.naef@obs.unige.ch}}

\date{Received / Accepted } 

\abstract{We report the detection of 3 new planetary companions 
orbiting the solar-type stars \object{{\footnotesize GJ}\,3021}, 
\object{{\footnotesize HD}\,52265} and 
\object{{\footnotesize HD}\,169830} using radial-velocity 
measurements taken with the {\footnotesize CORALIE} echelle 
spectrograph. All these planetary companions have longer orbital 
periods than the 51\,Peg-like objects. The orbits are fairly 
eccentric. The minimum masses of these planets range from 1 to 
3.3\,${\mathrm M_{\rm Jup}}$. The stars have spectral types from F8V 
to G6V. They are metal-rich. We also present our radial-velocity 
measurements for three solar-type stars known to host planetary 
companions \object{$\iota$\,Hor} (\object{{\footnotesize HD}\,17051}), 
\object{{\footnotesize HD}\,210277} and 
\object{{\footnotesize HD}\,217107}.
}

\titlerunning{The {\footnotesize CORALIE} survey for southern 
extrasolar planets V}
\authorrunning{D.~Naef et al.}
\maketitle

\section{Introduction}

Less than six years after the discovery of the first extrasolar 
planet orbiting the star \object{51\,Peg}  \citep{Mayor95}, the number 
of extrasolar planets detected around solar-type stars is now reaching 
63 (minimum masses below 10\,${\mathrm M_{\rm Jup}}$). None of the 
planetary companions detected so far resembles the giant planets of 
the solar system. The increase of the precision as well as the 
increase of the duration of the surveys make the radial-velocity 
searches sensitive to long period and very low mass companions. 
Sub-Saturnian planets have been recently detected 
\citep[ May the 4th 2000 {\footnotesize ESO PR}\footnote{www.eso.org/outreach/press-rel/pr-2000/pr-13-00.html}]{Marcy2000}. 
We find in the highlights of this research domain the detection of a 
3-planet system orbiting \object{$\upsilon$\,And} \citep{Butler99}, 
the detection of a 2-Saturnian planet system orbiting 
{\footnotesize HD\,83443} \citep{Mayorman} and the detection of the 
photometric \citep{Charbonneau,Henry2000} and spectroscopic 
\citep{trans} transits of the extrasolar planet orbiting the star 
\object{{\footnotesize HD}\,209458}. Planetary companions have also 
been detected in wide binary systems: see e.g. \object{16\,Cyg\,B} 
\citep{Cochran}, \object{{\footnotesize GJ}\,86} \citep{Quelozgj} or 
\object{{\footnotesize HD}\,80606} \citep{Naef2001}. It is interesting 
to notice as well that the majority of the stars with planets have a 
higher metal content than field stars in the solar vicinity
\citep{Gonz2000,Santosman,Santosmet,Santoslastmet}.

Some of the detected planets have short periods and very small 
orbital separations (the "Hot Jupiters"). Most of these Hot Jupiters 
are on circular orbits. There are also longer period objects. The 
orbital eccentricities of these objects almost cover the full 
possible range: from nearly circular (see e.g. the recently announced 
planet around {\footnotesize HD}\,28185, April the 4th  2001 
{\footnotesize ESO PR}\footnote{www.eso.org/outreach/press-rel/pr-2001/pr-07-01.html}) 
to nearly unity as in the case of {\footnotesize HD}\,80606 
\citep{Naef2001}. The 3 new extrasolar planet candidates we present 
in Sect.\,\ref{newplans} belong to this kind of objects 
(Jovian masses, longer periods, elongated orbits).

In this paper, we present the radial-velocity data obtained with 
{\footnotesize CORALIE} for the stars {\footnotesize GJ}\,3021, 
{\footnotesize HD}\,52265\footnote{The detection of the planetary 
companion around {\footnotesize HD}\,52265 was independently announced 
during the writing of this paper by \citet{Butler2000}} and 
{\footnotesize HD}\,169830. They reveal the presence of Jovian-mass 
companions. Section\,\ref{parentstars} describes the 
basic stellar properties of these stars. The Lithium abundances and 
the chromospheric activity of these objects are discussed in 
Sect.\,\ref{litab} and \ref{Ca}. The orbital solutions are presented 
in Sect.\,\ref{orbsolutions}. In Sect.\,\ref{confirms}, we present 
our radial-velocity measurements for the stars $\iota$\,Hor, 
{\footnotesize HD}\,210277 and {\footnotesize HD}\,217107.

\begin{table}[th!]
\caption{
\label{detections} Planetary and brown dwarf candidates detected with CORALIE}
\begin{tabular}{ll}
\hline
\noalign{\vspace{0.025cm}}
{\bf Star} & {\bf Reference} \\
\noalign{\vspace{0.025cm}}
\hline
\noalign{\vspace{0.025cm}}
\object{GJ\,86}                 & \citealt{Quelozgj}\\
\object{HD\,75289}              & \citealt{Udry2000}\\
\object{HD\,130322}             & \citealt{Udry2000}\\
\object{HD\,192263}             & \citealt{Santosposthd,Santoshd}\\
\object{GJ\,3021}               & \citealt{Naef2000}, this paper\\
\object{HD\,168746}             & Pepe et al. in prep.\\
\object{HD\,83443b}             & \citealt{Mayorman}\\
\object{HD\,108147}             & Pepe et al. in prep.\\
\object{HD\,52265}              & this paper\\
\object{HD\,82943b}             & Mayor et al. in prep.\\
\object{HD\,169830}             & this paper\\
\object{HD\,160202}$^{\dagger}$ & Udry et al. in prep\\
\object{HD\,202206}$^{\dagger}$ & Udry et al. in prep\\
\object{HD\,6434}               & \citealt{DQman}\\
\object{HD\,19994}              & \citealt{DQman}\\
\object{HD\,92788}$^{\ddagger}$ & \citealt{DQman}\\
\object{HD\,121504}             & \citealt{DQman}\\
\object{HD\,83443b}             & \citealt{Mayorman}\\
\object{HD\,168443c}            & \citealt{Udryman}\\
\object{HD\,28185}              & Santos et al. in prep.\\
\object{HD\,82943c}             & Mayor et al. in prep.\\
\object{HD\,141937}             & Udry et al. in prep.\\
\object{HD\,213240}             & Santos et al. in prep.\\
\noalign{\vspace{0.025cm}}
\hline
\noalign{\vspace{0.025cm}}
\end{tabular}
\\
$^{\dagger}$ Low-mass brown dwarf\\
$^{\ddagger}$ Independently announced by \citet{Fischer2001}
\end{table}

\section{Extrasolar planets orbiting the stars GJ\,3021, HD\,52265, and 
HD\,169830}\label{newplans}

\subsection{CORALIE high-precision radial-velocity survey}

Our radial-velocity measurements are done by computing the 
cross-correlation function ({\footnotesize CCF}) between the observed 
stellar spectrum and a numerical template. The instrumental velocity 
drifts are monitored and corrected using the "simultaneous Thorium 
referencing technique" with dual fibers 
\citep[more details in][]{Baranne96}. The achieved precision of this 
technique with {\footnotesize CORALIE} is currently of the order of 
7-8\,m\,s$^{-1}$ for almost three years \citep{DQanto}.
Three years after the start of the {\sl Geneva Southern Planet Search 
Programme} \citep{DQanto,Udryanto,Udry2000} at 
{\footnotesize ESO}-La\,Silla Observatory (Chile), 21 planetary and 2 
low-mass brown dwarf candidates have already been detected 
(see Table\,\ref{detections}).

The 298 {\footnotesize CORALIE} individual radial-velocity 
measurements presented in this paper are available in electronic form 
at the {\footnotesize CDS} in Strasbourg.

\subsection{Characteristics of the stars}\label{parentstars}

\subsubsection{GJ\,3021}\label{GJ3021star}

{\footnotesize GJ}\,3021 (\object{{\footnotesize HD}\,1237}, 
\object{{\footnotesize HIP}\,1292}) is a bright G6 dwarf located in 
the southern constellation of Hydrus. The astrometric parallax from 
the {\footnotesize HIPPARCOS} satellite \citep{ESA97} 
$\pi$\,=\,56.76\,$\pm$\,0.53\,mas sets the star at a distance of 
17.6\,pc from the Sun. The {\footnotesize HIPPARCOS} apparent 
magnitude and colour index are $m_{\rm V}$\,=\,6.59 and 
$B-V$\,=\,0.749. Thus the absolute magnitude is $M_{\rm V}$\,=\,5.36. 
With a bolometric correction $B.C.$\,=\,$-$0.16 \citep{Flower96}, we 
find an effective temperature of $T_{\rm eff}$\,=\,5417\,K and a 
luminosity of $L$\,=\,0.66\,L$_{\sun}$. Using 
{\footnotesize HIPPARCOS} proper motions and parallax and the 
$\gamma$\,velocity of {\footnotesize GJ}\,3021, we find 
$(U, V, W)$\,=\,(33.7, $-$17.4, 2.8)\,km\,s$^{-1}$. 
This velocity vector is similar to the one observed for stars in the 
Hyades Super-Cluster \citep{Chereul99}.

In a recent paper, \citet{Santosmet}, using a high signal-to-noise 
{\footnotesize CORALIE} spectrum, derived the following atmospheric 
parameters for {\footnotesize GJ}\,3021: 
$T_{\rm eff}$\,=\,5540\,$\pm$\,75\,K, $\log g$\,=\,4.7\,$\pm$\,0.2 
(cgs) and $[$Fe/H$]$\,=\,0.1\,$\pm$\,0.08.
From the {\footnotesize CORAVEL} {\footnotesize CCF} dip width, we 
obtain the projected rotational velocity 
$v\sin i$\,=\,5.5\,$\pm$\,1\,km\,s$^{-1}$ \citep{Benz84}.\label{vsini} 
The calibration by \citet{Benz84} does not account for metallicity 
effects. In their calibration, the intrinsic line width for 
non-rotating stars is supposed to be mainly a function of the 
$B-V$ colour index (pressure broadening by Van der Waals effect). 
For metal-rich stars, the number of saturated lines increases. This 
causes the CCF intrinsic width to increase. With the calibration by 
\citet{Benz84}, the intrinsic line width is therefore under-estimated 
and the rotational broadening over-estimated for metal-rich stars. 
The $v\sin i$ value indicated above is thus an upper-limit. 
The \mbox{{\footnotesize ROSAT All-Sky Survey}} X--ray luminosity for 
{\footnotesize GJ}\,3021 is 
$L_{\rm X}$\,=\,103.6$\cdot$$10^{27}$\,erg\,$s^{-1}$ \citep{Hunsch99}.

{\footnotesize GJ}\,3021 is not photometrically stable. The scatter of 
the {\footnotesize HIPPARCOS} photometric measurements is of the 
order of 12\,mmag. In the {\footnotesize HIPPARCOS} catalogue, 
{\footnotesize GJ}\,3021 is classified as a possible micro-variable 
star. The Fourier transform of the photometric data shows no clear 
significant frequency. It is not surprising, assuming that the 
photometric signal is due to features such as spots on the 
stellar surface. The typical lifetimes of spots are known to be much 
shorter than the duration of the {\footnotesize HIPPARCOS} mission.

\begin{table*}[th!]
\caption{
\label{allstarparam}
Observed and inferred stellar parameters for GJ\,3021, HD\,52265 and 
HD\,169830. The spectral types, apparent magnitudes, color indexes, 
parallaxes and proper motions are from HIPPARCOS \citep{ESA97}. The 
bolometric correction is computed with the calibration by 
\citet{Flower96}. The atmospheric parameters $T_{\rm eff}$, $\log g$, 
$[$Fe/H$]$ and $\xi _{\rm t}$ are from \citet{Santosmet}. 
The X--ray luminosities are from \citet{Hunsch99} for GJ\,3021 and from 
\citet{Piters} for HD\,169830. The projected velocities are derived 
from CORAVEL with the calibration by \citet{Benz84}. The ages and 
masses of HD\,52265 and HD\,169830 are derived from the Geneva 
evolutionary tracks \citep{Schaerer93}}
\begin{tabular}{ll|r@{  $\,\pm\,$  }lr@{  $\,\pm\,$  }lr@{  $\,\pm\,$  }l}
\hline
\noalign{\vspace{0.025cm}}
{\bf Parameter}             & {\bf Units}                & \multicolumn{2}{c}{\bf {\footnotesize GJ}\,3021}  & \multicolumn{2}{c}{\bf {\footnotesize HD}\,52265} & \multicolumn{2}{c}{\bf {\footnotesize HD}\,169830}\\
\noalign{\vspace{0.025cm}}
\hline
\noalign{\vspace{0.025cm}}
$Sp.\,Type$                 &                            & \multicolumn{2}{c}{G6V}                           & \multicolumn{2}{c}{G0V}                           & \multicolumn{2}{c}{F8V}          \\
$m_{\rm V}$                 &                            & \multicolumn{2}{c}{6.59}                          & \multicolumn{2}{c}{6.29}                          & \multicolumn{2}{c}{5.90}         \\
$B-V$                       &                            & 0.749    & 0.004                                  & 0.572     & 0.003                                 & 0.517     & 0.004                \\
$\pi$                       & (mas)                      & 56.76    & 0.53                                   & 35.63     & 0.84                                  & 27.53     & 0.91                 \\
$Distance$                  & (pc)                       & 17.62    & $^{0.17}_{0.16}$                       & 28.07     & $^{0.68}_{0.65}$                      & 36.32     & $^{1.24}_{1.16}$     \\
$\mu _{\alpha}\cos(\delta)$ & (mas yr$^{-1}$)            & 433.87   & 0.53                                   & $-$115.76 & 0.68                                  & $-$0.84   & 1.23                 \\
$\mu _{\delta}$             & (mas yr$^{-1}$)            & $-$57.95 & 0.48                                   & 80.35     & 0.55                                  & 15.16     & 0.72                 \\
$M_{\rm V}$                 &                            & \multicolumn{2}{c}{5.36}                          & \multicolumn{2}{c}{4.05}                          & \multicolumn{2}{c}{3.10}         \\
$B.C.$                      &                            & \multicolumn{2}{c}{$-$0.16}                       & \multicolumn{2}{c}{$-$0.046}                      & \multicolumn{2}{c}{$-$0.02}      \\
$L$                         & (L$_{\sun}$)               & \multicolumn{2}{c}{0.66}                          & \multicolumn{2}{c}{1.98}                          & \multicolumn{2}{c}{4.63}         \\
$T_{\rm eff}$               & ($\degr$K)                 & 5540     & 75                                     & 6060     & 50                                     & 6300     & 50                    \\
$\log g$                    & (cgs)                      & 4.70     & 0.20                                   & 4.29     & 0.25                                   & 4.11     & 0.25                  \\
$\xi _{\rm t}$              & (km\,s$^{-1}$)             & 1.47     & 0.10                                   & 1.29     & 0.10                                   & 1.37     & 0.10                  \\
$[$Fe/H$]$                  &                            & 0.10     & 0.08                                   & 0.21     & 0.06                                   & 0.21     & 0.05                  \\ 
$M_{\rm 1}$                 & (M$_{\sun}$)               & \multicolumn{2}{c}{0.9}                           & \multicolumn{2}{c}{1.18}                          & \multicolumn{2}{c}{1.4}          \\
$R$                         & (R$_{\sun}$)               & \multicolumn{2}{c}{0.9}                           & \multicolumn{2}{c}{1.1}                           & \multicolumn{2}{c}{1.2}          \\
$L_{\rm X}$                 & ($10^{27}$\,erg\,s$^{-1}$) & \multicolumn{2}{c}{103.6}                         & \multicolumn{2}{c}{...}                           & \multicolumn{2}{c}{$<$\,398}     \\
$v\sin i$$^{\dagger}$       & (km\,s$^{-1}$)             & 5.5      & 1                                      & 5.2      & 1                                      & 3.8      & 1                     \\
$Age$                       & (Gyr)                      & \multicolumn{2}{c}{...}                           & \multicolumn{2}{c}{3.5}                           & \multicolumn{2}{c}{2.3}          \\
\noalign{\vspace{0.025cm}}
\hline
\noalign{\vspace{0.025cm}}
\end{tabular}
\\
$^{\dagger}$ over-estimated for metal-rich stars, see Sect.\,\ref{GJ3021star}
\end{table*}

Finally, no infrared excess was detected for {\footnotesize GJ}\,3021 
at 60\,$\mu$m \citep[{\footnotesize ISO} data,][]{Decin}. A 
{\footnotesize CORALIE} radial-velocity monitoring of a set of G 
dwarfs observed in the infrared by \citet{Decin} is on its way. The 
main goal of this combined survey is to study the correlation 
(or anti-correlation) between the presence of an infrared excess and 
the occurrence of planets or companion stars. In the selected sample, 
two stars actually have a planetary companion, 
{\footnotesize GJ}\,3021 (this paper) and {\footnotesize GJ}\,86 
\citep{Quelozgj}, without trace of infrared excess.

Observed and inferred stellar parameters are summarized in 
Table\,\ref{allstarparam} with references to the corresponding 
observations or methods used to derive the quantities.

\subsubsection{HD\,52265}

{\footnotesize HD}\,52265 (\object{{\footnotesize HIP}\,33719}) is a 
bright ($m_{\rm V}$\,=\,6.29) G0 dwarf located in the constellation 
of Monoceros (the Unicorn) 28.1\,pc away from the Sun. 
Using the {\footnotesize HIPPARCOS} \citep{ESA97} precise astrometric 
parallax ($\pi$\,=\,35.63\,$\pm$\,0.84\,mas), we infer an absolute 
magnitude of $M_{\rm V}$\,=\,4.05. This value, slightly above the 
main sequence, disagrees with the commonly assumed spectral 
classification G0\,III-IV for this star. With the bolometric 
correction ($B.C.$\,=\,$-$0.046) and temperature scale calibrated by 
\citet{Flower96}, we find an effective temperature 
$T_{\rm eff}$\,=\,5995\,K and a luminosity $L$\,=\,1.98\,L$_{\sun}$.

With a high signal-to-noise {\footnotesize CORALIE} spectrum, 
\citet{Santosmet} derived the following stellar atmospheric 
parameters for {\footnotesize HD}\,52265: 
$T_{\rm eff}$\,=\,6060\,$\pm$\,50\,K, $\log g$\,=\,4.29\,$\pm$\,0.25 
(cgs) and $[$Fe/H$]$\,=\,0.21\,$\pm$\,0.06. Using the Geneva stellar 
evolution models \citep{Schaerer93} with $T_{\rm eff}$\,=\,6060\,K, 
$M_{\rm V}$\,=\,4.05 and $[$Fe/H$]$\,=\,0.21, we measure a mass 
$M_{\rm 1}$\,=\,1.18\,M$_{\sun}$ and an age  of 3.5\,Gyr.
{\footnotesize HD}\,52265 appears to be a slightly evolved metal-rich 
G0 dwarf.

From {\footnotesize CORAVEL} measurements, we compute 
$v\sin i$\,=\,5.2\,$\pm$\,1\,km\,s$^{-1}$ 
\citep[ over-estimated, see Sect.\,\ref{GJ3021star}]{Benz84}.
Finally, {\footnotesize HD}\,52265 is photometrically stable 
\citep[{\footnotesize HIPPARCOS} photometry, ][]{ESA97}. The main 
stellar characteristics of {\footnotesize HD}\,52265 are summarized in 
Table\,\ref{allstarparam}.

\subsubsection{HD\,169830}

{\footnotesize HD}\,169830 (\object{HIP\,90485}) is a bright 
($m_{\rm V}$\,=\,5.9) F8 dwarf located in the constellation of 
Sagittarius (the Archer) 36.3\,pc away from the Sun. With the 
{\footnotesize HIPPARCOS} \citep{ESA97} astrometric parallax 
($\pi$\,=\,27.53\,$\pm$\,0.91\,mas), the bolometric correction 
($B.C.$\,=\,$-$0.19) and the temperature scale calibrated by 
\citet{Flower96}, we compute the absolute magnitude, the effective 
temperature and the luminosity of {\footnotesize HD}\,169830: 
$M_{\rm V}$\,=\,3.10, $T_{\rm eff}$\,=\,6211\,K and 
$L$\,=\,4.63\,L$_{\sun}$. Using Str\"omgren photometry, \citet{Edv93} 
have inferred stellar atmospheric parameters for 
{\footnotesize HD}\,169830. They found $T_{\rm eff}$\,=\,6382\,K, 
$\log g$\,=\,4.15 (cgs), $[$M/H$]$\,=\,0.17 and $[$Fe/H$]$\,=\,0.13. 
With a high signal-to-noise {\footnotesize CORALIE} spectrum, 
\citet{Santosmet} derived similar values for the atmospheric 
parameters ($T_{\rm eff}$\,=\,6300\,$\pm$\,50\,K, 
$\log g$\,=\,4.11\,$\pm$\,0.25 (cgs) and 
$[$Fe/H$]$\,=\,0.21\,$\pm$\,0.05). {\footnotesize HD}\,169830 is a 
slow rotator. Using {\footnotesize CORAVEL} data, we have 
$v\sin i$\,=\,3.8\,$\pm$\,1\,km\,s$^{-1}$ 
\citep[ over-estimated, see Sect.\,\ref{GJ3021star}]{Benz84}.
\citet{Edv93} also estimated the age of {\footnotesize HD}\,169830 
using its position in a colour-magnitude diagram. They found an age of 
2.2\,Gyr. This value is in agreement with the 2\,Gyr found by 
\citet{Ng} and also consistent with the value obtained using the 
Geneva evolutionary tracks \citep{Schaerer93}: 
$age$\,=\,2.3\,Gyr, $M_{\rm 1}$\,=\,1.4\,M$_{\sun}$.

Using data from the \mbox{{\footnotesize ROSAT All-Sky Survey}}, 
\citet{Piters} obtained an upper limit for the X--ray luminosity of 
{\footnotesize HD}\,169830: 
$L_{\rm X}$\,$<$\,398$\cdot$$10^{27}$\,erg\,s$^{-1}$. A carefull 
analysis of the {\footnotesize ROSAT} data (M. Audard, priv. comm.) 
shows that there is no source at the position of the star. The X--ray 
flux value is consistant with the background noise. Note that there 
is a bright nearby X--ray source (the {\footnotesize LMXB} Sgr X--4) 
that could be responsible for the high X--ray background.
The main stellar characteristics of {\footnotesize HD}\,169830 are 
summarized in Table\,\ref{allstarparam}.

\begin{figure*}[th!]
    \psfig{width=\hsize,file=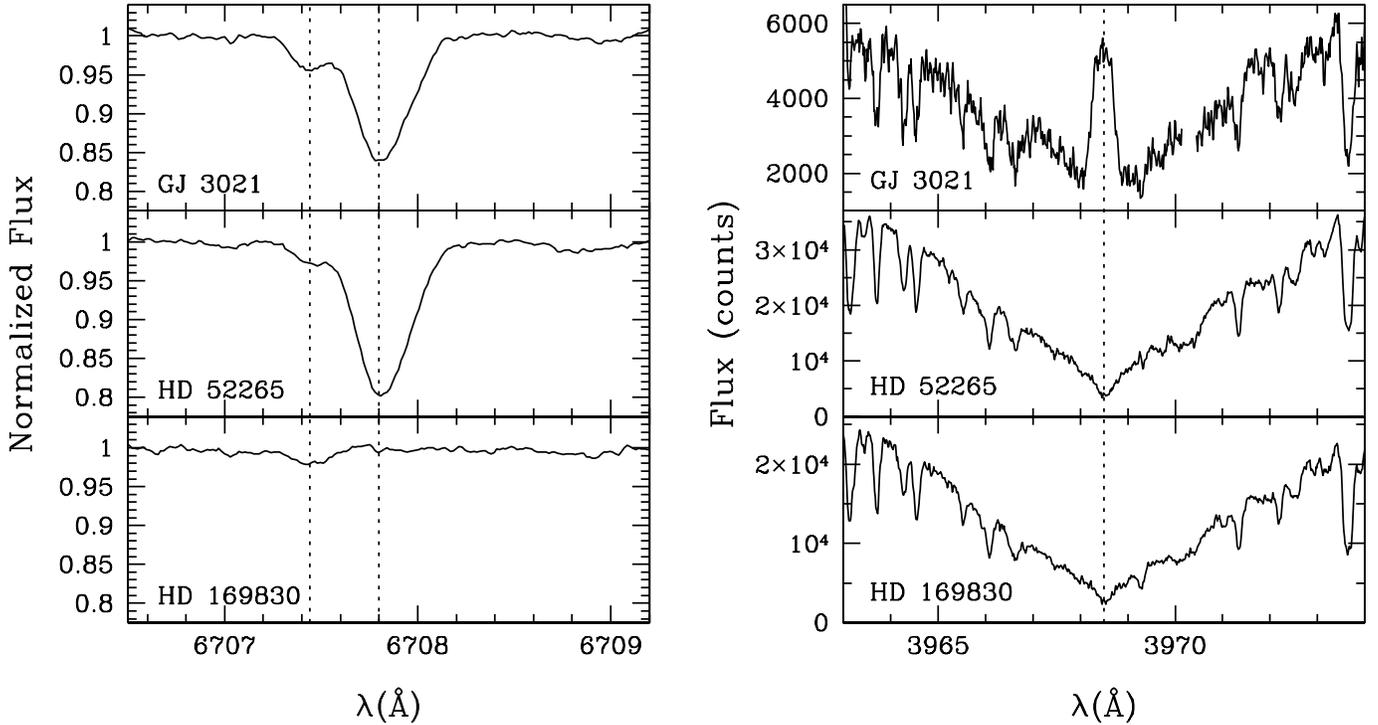}
    \caption{
    \label{litcalallplots}
    Results of the addition of the CORALIE spectra. 
    Left: The $\lambda$\,6707.8\,\AA\,\ion{Li}{i} absorption line 
    region for GJ\,3021, HD\,52265 and HD\,169830. The Lithium is 
    clearly detected for GJ\,3021 and HD\,52265. The dotted lines 
    indicate the positions of the $\lambda$\,6707.4\,\AA\,\ion{Fe}{i} 
    and $\lambda$\,6707.8\,\AA\,\ion{Li}{i} lines. 
    Right: the $\lambda$\,3968.5\,\AA\,\ion{Ca}{ii}\,H absorption line 
    region. A strong chromospheric emission is detected for GJ\,3021. 
    The two other candidates do not exhibit any trace of chromospheric 
    emission}
\end{figure*}

\subsection{Lithium abundances} \label{litab}

In this section, we estimate the Lithium abundance of the host stars 
with {\footnotesize CORALIE} spectra. In order to obtain a high 
signal in the \mbox{$\lambda$ 6707.8 \AA\ \ion{Li}{i}} line region, 
we added all the measured spectra. The latter have been measured with 
our standard instrumental configuration. In this configuration, the 
Thorium-Argon lamp is used, adding a reference spectrum on the 
{\footnotesize CCD}. This reference spectrum is located between the 
orders of the stellar echelle spectrum. This configuration, needed for 
high-precision radial-velocity measurements, is not optimal for 
spectroscopic determinations (such as abundance determinations). 
However the spectra measured in this configuration are usable for 
limited spectroscopic determinations. The main interest of such 
determinations is the possibility of getting spectroscopic 
informations for very large stellar samples without specific 
measurements (i.e. without the need of additional observing time).

The two difficulties of this configuration for abundance 
determinations are: the additional background contamination and the 
shorter distance between the orders preventing us from correctly 
estimating the background contamination level. The measured fluxes 
are then not corrected for the background contamination effect. The 
obtained spectral lines have weaker contrasts and therefore smaller 
equivalent widths. The typical background contamination level measured 
for {\footnotesize ELODIE} \citep{Baranne96}, a similar instrument, is 
about \mbox{3 \%} in the \mbox{$\lambda$ 6707.8 \AA\ \ion{Li}{i}} line 
region. We have used this value to correct our equivalent width 
measurements.

We also used two {\footnotesize GJ}\,3021 spectra taken without 
calibration lamp (single fiber mode) in order to check our correction 
scheme. In this case, the background contamination is corrected on the 
{\footnotesize CCD} frame. The corrected equivalent width measured 
with the calibration lamp (57.7\,m\AA) is in good agreement with the 
value measured without calibration lamp (57.6\,m\AA).

\begin{table}[th!]
\caption{
\label{litfit} 
Results of double-gaussian fits to the \ion{Fe}{i}-\ion{Li}{i} blend 
and Lithium abundances for GJ\,3021 and HD\,52265. 
$W_{\lambda,{\rm cor}}$ is the equivalent width corrected for the 
assumed 3\,\% background contamination level. The values in 
parentheses after the abundance values indicate the change in 
$\log$\,$n({\rm Li})$ resulting from a 1$\sigma$ change in the 
effective temperature without accounting for other error sources. 
More realistic errors are probably not better than 0.1-0.15 dex. 
Note the good agreement between our abundances and the values by 
Israelian et al. (in prep.)}
\begin{tabular}{ll|cc}
\hline
\noalign{\vspace{0.025cm}}
&                                        & {\bf {\footnotesize GJ}\,3021} & {\bf {\footnotesize HD}\,52265}\\
\noalign{\vspace{0.025cm}}
\hline
\noalign{\vspace{0.025cm}}
& &\multicolumn{2}{c}{\ion{Fe}{i} Line} \\
\noalign{\vspace{0.025cm}}
\hline
\noalign{\vspace{0.025cm}}
$\lambda$                         & \AA  & 6707.438                       & 6707.441        \\
$W_{\lambda,{\rm cor}}$           & m\AA & 8.38                           & 3.97            \\
\noalign{\vspace{0.025cm}}
\hline
\noalign{\vspace{0.025cm}}
& &\multicolumn{2}{c}{\ion{Li}{i} Line}\\
\noalign{\vspace{0.025cm}}
\hline
\noalign{\vspace{0.025cm}}
$\lambda$                         & \AA  & 6707.832                       & 6707.837        \\
$W_{\lambda,{\rm cor}}$           & m\AA & 57.56                          & 69.33           \\
$T_{\rm eff}$	                  & K    & 5540\,$\pm$\,75                & 6060\,$\pm$\,50 \\
$\log$ $W_{\lambda,{\rm cor}}$    & m\AA & 1.76                           & 1.84            \\
$\log$\,$n({\rm Li})$             &      & 2.11 (0.08)                    & 2.70 (0.04)     \\
$\log$\,$n({\rm Li})$$^{\dagger}$ &      & 2.11                           & 2.77            \\
\noalign{\vspace{0.025cm}}
\hline
\noalign{\vspace{0.025cm}}
\end{tabular}

$^{\dagger}$ Israelian et al. (in prep.)\\
\end{table}

Figure\,\ref{litcalallplots} (left) shows the spectra we obtain after 
adding all the available spectra of {\footnotesize GJ}\,3021, 
{\footnotesize HD}\,52265 and {\footnotesize HD}\,169830. The 
signal-to-noise ratio of these spectra is of the order of 300 at the 
continuum. The Lithium absorption feature is clearly visible for 
{\footnotesize GJ}\,3021 and {\footnotesize HD}\,52265. There are no 
clear traces of Lithium for {\footnotesize HD}\,169830. The 
\mbox{$\lambda$ 6707.4 \AA\ \ion{Fe}{i}} absorption feature is clearly 
visible for the 3 stars on the blue edge of the Lithium line.
Table\,\ref{litfit} shows the results of double gaussian fits to the 
\ion{Fe}{i}--\ion{Li}{i} blend for {\footnotesize GJ}\,3021 and 
{\footnotesize HD}\,52265.

To estimate the Lithium abundance, we have used the curves of growth 
tabulated by \citet{Soder93} computed for the Pleiades metallicity. 
The Lithium abundances then obtained from our equivalent widths are 
listed in Table\,\ref{litfit} (on a scale where 
$\log$\,$n({\rm H})$\,=\,12). Our determination of the Lithium 
abundance for {\footnotesize GJ}\,3021 corresponds to the typical 
Lithium abundance measured for G6 dwarfs in the Hyades 
\citep[see e.g.][]{Soder93}. For {\footnotesize HD}\,169830, an 
upper limit of the Lithium equivalent width can be estimated. We have 
(2$\sigma$ confidence level): 
$W_{\lambda}$\,$<$\,2\,(S/N)$^{-1}$\,$\Sigma$\,$\sqrt{2\pi}$ where 
$\Sigma$ is a typical line width. With S/N\,=\,300 and 
$\Sigma$\,=\,130\,m\AA, we have: $W_{\lambda}$\,$<$\,2.17\,m\AA. 
Correcting this value for the background contamination, we have 
$W_{\lambda,{\rm max}}$\,=2.24\,m\AA. This value provides the 
following constraint on the Lithium abundance of 
{\footnotesize HD}\,169830: $\log$\,$n({\rm Li})$\,$<$\,1.31 
(2$\sigma$ confidence level).

Our determinations agree with a recent study by Israelian et al. 
(in prep.). The values they obtained for {\footnotesize GJ}\,3021 
and {\footnotesize HD}\,52265 are respectively  
$\log$\,$n({\rm Li})$\,=\,2.11 and 2.77. They also obtained an upper 
limit for the Lithium abundance of {\footnotesize HD}\,169830: 
$\log$\,$n({\rm Li})$\,$<$\,1.12.

\subsection{Chromospheric activity}\label{Ca}

Figure\,\ref{litcalallplots} (right) shows the 
\mbox{$\lambda$ 3968.5 \AA\ \ion{Ca}{ii} H} absorption line region for 
{\footnotesize GJ}\,3021, {\footnotesize HD}\,52265 and 
{\footnotesize HD}\,169830. These spectra were obtained by adding all 
our {\footnotesize CORALIE} spectra. We have derived the 
$\log (R^{\prime}_{\rm HK})$ chromospheric activity indicator for the 
three primary stars (\citealp[see][]{Santospost,Santosact} for details 
about the technique). Table\,\ref{lrphkall} shows the results: 
{\footnotesize HD}\,52265 and {\footnotesize HD}\,169830 are 
chromospherically inactive and {\footnotesize GJ}\,3021 is very active. 
For the latter, we find $\log (R^{\prime}_{\rm HK})$\,=\,$-$4.27, 
a value substantially higher than the one by 
\citet[ $\log (R^{\prime}_{\rm HK})$\,=\,$-$4.44]{HenryT96} whose 
result is based on only one measurement without estimation of the 
scatter. We believe that given our estimate of the scatter, the 
discrepancy between the two values is due to the variability of the 
activity level of {\footnotesize GJ}\,3021. 

With the $\log (R^{\prime}_{\rm HK})$ value, we can estimate the 
rotational periods (using the calibration in \citealt{Noyes84}) and 
the stellar ages (using the calibration in \citealt{Donahue93} also 
quoted in \citealt{HenryT96}). The rotational periods we obtain  for 
{\footnotesize HD}\,52265 and {\footnotesize HD}\,169830 are 14.6 and 
9.5\,days, respectively and ages of 4\,Gyr for both stars. 
For {\footnotesize GJ}\,3021, the scatter of our activity 
measurements is large (see Table\,\ref{lrphkall}). With our mean 
activity value $\log (R^{\prime}_{\rm HK})$\,=\,$-$4.27, we derive a 
rotational period of 4.1\,days and an age of 0.02\,Gyr. Estimates of 
the maximum rotational period and age can be obtained by using our 
lowest activity value $\log (R^{\prime}_{\rm HK})$\,=\,$-$4.49 
yielding $P_{\rm rot,max}$\,=\,12.6\,days and 
$Age_{\rm max}$\,=\,0.8\,Gyr. This maximum age estimate is compatible 
with the age of the Hyades.

\begin{table}
\caption{
\label{lrphkall} 
Measured chromospheric activity for the three primary stars using 
CORALIE spectra. $N_{\rm meas}$ indicates the number of usable spectra. 
$S_{\rm COR}$ is the activity index obtained with CORALIE. Activity 
values have been calibrated to the Mount-Wilson system \citep{Vaughan} 
before computing the $\log (R^{\prime}_{\rm HK})$ indicator so that it 
can be compared to the values by \citet{HenryT96}. Notice the high 
$S_{\rm COR}$ derived for GJ\,3021\ and the large dispersion around 
this mean value}
\begin{tabular}{lcccc}
\hline
\noalign{\vspace{0.025cm}}
{\bf Star}                 & {\bf $N_{\rm meas}$} & {\bf $S_{\rm COR}$} & {\bf $\sigma$($S_{\rm COR}$)} & {\bf $\log (R^{\prime}_{\rm HK})$} \\
\noalign{\vspace{0.025cm}}
\hline
\noalign{\vspace{0.025cm}}
{\footnotesize GJ}\,3021   & 20                   & 0.578               & 0.11                          & $-$4.27\\
{\footnotesize HD}\,52265  & 31                   & 0.146               & 0.03                          & $-$4.91\\
{\footnotesize HD}\,169830 & 11                   & 0.136               & 0.02                          & $-$4.93\\
\noalign{\vspace{0.025cm}}
\hline
\end{tabular}
\end{table}

\begin{table*}[th!]
\caption{\label{allorbs_par} 
Fitted orbital elements for GJ\,3021, HD\,52265 and HD\,169830 and 
characteristics of their planetary companions}
\begin{tabular}{lc|r@{  $\,\pm\,$  }lr@{  $\,\pm\,$  }lr@{  $\,\pm\,$  }l}
\hline
\noalign{\vspace{0.025cm}}
{\bf Parameter}    & {\bf Units}                                             & \multicolumn{2}{c}{\bf {\footnotesize GJ}\,3021} &  \multicolumn{2}{c}{\bf {\footnotesize HD}\,52265} & \multicolumn{2}{c}{\bf {\footnotesize HD}\,169830}\\
\noalign{\vspace{0.025cm}}
\hline
\noalign{\vspace{0.025cm}}
$P$                                & (days)                                  & 133.71     & 0.20                                & 119.6     & 0.42                                   & 229.9     & 4.6             \\
$T$                                & ({\footnotesize HJD}\,$-$\,2\,400\,000) & 51\,545.86 & 0.64                                & 51\,422.3 & 1.7                                    & 51\,485.9 & 4.8             \\
$e$                                &                                         & 0.511      & 0.017                               & 0.35      & 0.03                                   & 0.35      & 0.04            \\
$\gamma$                           & (km\,s$^{-1}$)                          & $-$5.806   & 0.003                               & 53.769    & 0.001                                  & $-$17.215 & 0.003           \\
$\omega$                           & ($\degr$)                               & 290.7      & 3.0                                 & 211       & 6                                      & 183       & 7               \\
$K_{\rm 1}$                        & (m\,s$^{-1}$)                           & 167        & 4                                   & 42        & 1                                      & 83        & 4               \\
$a_{\rm 1} \sin i$                 & ($10^{-4}$AU)                           & 17.6       & 0.4                                 & 4.26      & 0.15                                   & 16.4      & 0.7             \\
$f_{\rm 1}(m)$                     & ($10^{-10}$M$_{\odot}$)                 & 412        & 31                                  & 7.33      & 0.08                                   & 113       & 14              \\
\noalign{\vspace{0.025cm}}
\hline
\noalign{\vspace{0.025cm}}
$N_{\rm meas}$                     &                                         & \multicolumn{2}{c}{61}                           & \multicolumn{2}{c}{71}                             & \multicolumn{2}{c}{35}\\
$<\epsilon_{\rm RV}>$$^{\ddagger}$ & (m\,s$^{-1}$)                           & \multicolumn{2}{c}{8.2}                          & \multicolumn{2}{c}{5.6}                            & \multicolumn{2}{c}{5.5}\\
$\sigma(O-C)$                      & (m\,s$^{-1}$)                           & \multicolumn{2}{c}{19.2}                         & \multicolumn{2}{c}{7.3}                            & \multicolumn{2}{c}{9.2}\\
$\chi^{\rm 2}$                     &                                         & \multicolumn{2}{c}{188.18}                       & \multicolumn{2}{c}{45.52}                          & \multicolumn{2}{c}{31.45}\\
P($\chi^{\rm 2}$)                  &				             & \multicolumn{2}{c}{$<$\,5$\cdot$10$^{-7}$}       & \multicolumn{2}{c}{0.968}                          & \multicolumn{2}{c}{0.344}\\
\noalign{\vspace{0.025cm}}
\hline
\noalign{\vspace{0.025cm}}
$m_{\rm 2}\sin i$                  & (M$_{\rm Jup}$)                         & 3.37           & 0.09                            & 1.05          & 0.03                               & 2.94          & 0.12\\
$a$                                & (AU)                                    & \multicolumn{2}{c}{0.49}                         & \multicolumn{2}{c}{0.5}                            & \multicolumn{2}{c}{0.82}\\
$d_{\rm Periastron}$               & (AU)                                    & \multicolumn{2}{c}{0.25}                         & \multicolumn{2}{c}{0.325}                          & \multicolumn{2}{c}{0.54}\\
$d_{\rm Apastron}$                 & (AU)                                    & \multicolumn{2}{c}{0.75}                         & \multicolumn{2}{c}{0.675}                          & \multicolumn{2}{c}{1.1}\\
$T_{\rm equ}$$^\natural$           & ($\degr$K)                              & \multicolumn{2}{c}{260--440}                     & \multicolumn{2}{c}{330--480}                       & \multicolumn{2}{c}{280--410} \\
\noalign{\vspace{0.025cm}}
\hline
\noalign{\vspace{0.025cm}}
\end{tabular}
\\
$^\ddagger$ Mean photon-noise error,
$^\natural$ Equilibrium temperature, \citet{Guillot96}
\end{table*}

\subsection{Orbital solutions}\label{orbsolutions}\label{gj3021orbsol}

\begin{figure*}[th!]
    \psfig{file=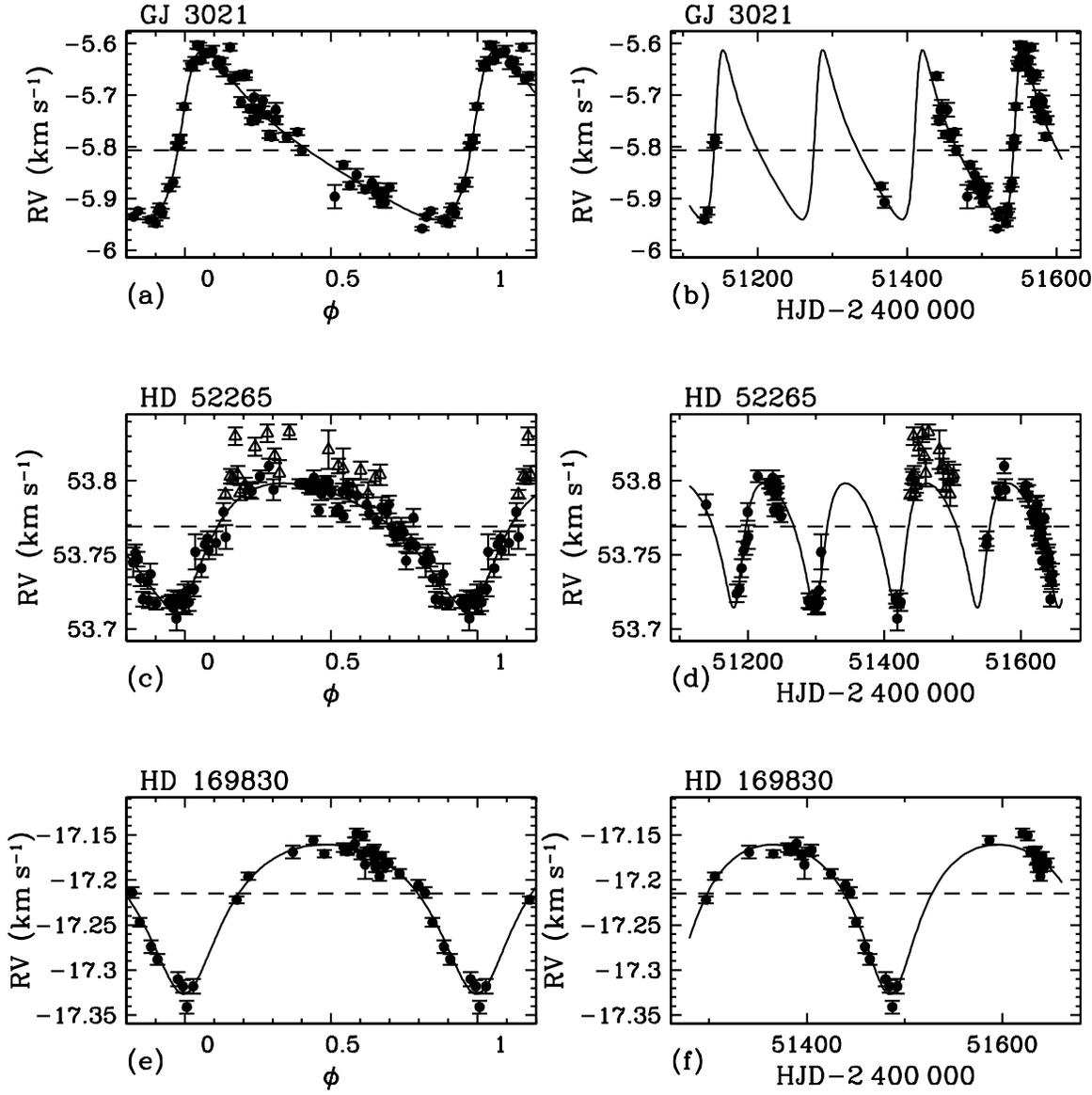,width=0.85\hsize}
    \caption{
    \label{allorbs}
    CORALIE data: fitted orbital solutions for GJ\,3021, HD\,52265 and 
    HD\,169830. Left column: phase-folded velocities. Right Column: 
    temporal velocities. The open triangles in panels {\bf (c)} and 
    {\bf (d)} indicate measurements that have not been used for 
    computing the orbital solution of HD\,52265}
\end{figure*}

61 and 35 {\footnotesize CORALIE} high precision radial-velocity 
measurements are available for {\footnotesize GJ}\,3021 and 
{\footnotesize HD}\,169830, respectively. These measurements were 
obtained between {\footnotesize HJD}\,=\,2\,451\,128 and 
{\footnotesize HJD}\,=\,2\,451\,590 for {\footnotesize GJ}\,3021 and 
between {\footnotesize HJD}\,=\,2\,451\,296 and 
{\footnotesize HJD}\,=\,2\,451\,668 for {\footnotesize HD}\,169830. 
Table\,\ref{allorbs_par} provides the fitted orbital elements to these 
measurements as well as the computed minimum masses of the planetary 
companions and their semi-major axes. The radial-velocity are 
displayed in Fig.\,\ref{allorbs}.

91 {\footnotesize CORALIE} radial-velocity measurements of 
{\footnotesize HD}\,52265 have been gathered between 
{\footnotesize HJD}\,=\,2\,451\,138 and 
{\footnotesize HJD}\,=\,2\,451\,647. The mean photon-noise error of 
these velocities is 5.6\,\,m\,s$^{-1}$. During the first observing 
season, we found the orbital solution but we decided to wait for 
additional measurements to check the validity of this solution. The 
strange behaviour of some of the measurements of the second season 
pushed us to delay the announcement of the detection 
(see open triangles in Figs.\,\ref{allorbs}c and \ref{allorbs}d). 
Up to now, we still have no explanation for what happened between 
{\footnotesize HJD}\,=\,2\,451\,430 and 
{\footnotesize HJD}\,=\,2\,451\,500. We did not find this kind of 
behaviour on any other star in our sample and no instrumental problem 
was reported during this period. We decided at the end not to use 
these measurements to fit the orbital solution and only 71 
radial-velocity measurements have been finally used.

\subsection{GJ\,3021 radial-velocity residuals and X--ray emission}

The residuals from the fitted orbit for {\footnotesize GJ}\,3021 are 
abnormally large (P($\chi^{\rm 2}$)\,$<$\,5$\cdot$10$^{-7}$, see 
Table\,\ref{allorbs_par}). With a mean photon-noise uncertainty of 
8.2\,m\,s$^{-1}$\ and an assumed systematic error of 7\,m\,s$^{-1}$, 
the expected value is of the order of 10-11\,m\,s$^{-1}$. The 
extra-scatter is therefore of the order of 15-16\,m\,s$^{-1}$. 
{\footnotesize GJ}\,3021 is an active star (see Sect.\,\ref{Ca}). It 
is well known \citep{Saar98,Santosact,Santospost,Santosbisec} that 
features on the surface of active stars (such as spots) induce 
radial-velocity variations ("jitter") due to line profile variations. 
The predicted jitter for a very active G dwarf is of the order of 
15-20\,m\,s$^{-1}$ \citep{Santosact,Santospost,Santosbisec}. We have 
computed the bisectors of our observed {\footnotesize CCFs} 
\citep[for detail, see][]{Queloz2001} for {\footnotesize GJ}\,3021. 
The bisector span is variable ($P(\chi^{2})$\,$<$\,0.001). The 
dispersion of the bisector span measurements is larger 
($\simeq$\,24\,m\,s$^{-1}$) than the observed extra-scatter 
($\simeq$\,15\,m\,s$^{-1}$). A weak correlation between the bisector 
span and the observed residuals is detected. The photon-noise errors 
affecting our span measurements (mean error value: 
$\simeq$\,16\,m\,s$^{-1}$) are too high to enable us to clearly 
establish this correlation. At this point, we cannot conclude 
that the abnormal radial-velocity residuals result from line profile 
variations for {\footnotesize GJ}\,3021.

A correlation between rotation and X--ray emission is known to 
exist \citep[see e.g][]{Queloz98}. This correlation can be well 
illustrated using a $\log(L_{\rm X}/L_{\rm Bol})$ versus 
$\log(P_{\rm rot}/\tau_{\rm c})$\,$-$\,$\log(\sin i)$ diagram 
(Rossby diagram). For {\footnotesize GJ}\,3021, we have 
$\log(L_{\rm X}/L_{\rm Bol})$\,$\simeq$\,$-$4.4. Using our measured 
$v\sin i$, $\log(\tau_{\rm c})$\,=\,1.238 \citep[from the calibration 
in][]{Noyes84} and assuming a stellar radius of 
$R$\,=\,0.9\,R$_{\odot}$, we find 
$\log(P_{\rm rot}/\tau_{\rm c})$\,$-$\,$\log(\sin i)$\,=\,$-$0.320. 
This value corresponds to the value expected for this 
$\log(L_{\rm X}/L_{\rm Bol})$ ratio \citep[see Fig.\,9 in][]{Queloz98}. The 
X--ray emission of {\footnotesize GJ}\,3021 probably results from 
activity-related processes.

\section{The planetary companions around $\iota$\,Hor, HD\,210277 and 
HD\,217107}\label{confirms}

In this section, we present the {\footnotesize CORALIE} measurements 
and orbital solutions for the solar-type stars $\iota$\,Hor, 
{\footnotesize HD}\,210277 and {\footnotesize HD}\,217107. 

\subsection{The planetary companion of $\iota$\,Hor}

Indications of the potential presence of a planetary companion around 
the young solar-type star $\iota$\,Hor\ 
(\object{{\footnotesize HD}\,17051}, 
\object{{\footnotesize HIP}\,12653}) were first presented during the 
summer 1998 by \citet{Kurster98,Kurster99}. During the summer 1999, 
these authors have published their final solution for the planetary 
companion \citep{Kurster2000}: a 320--day non-circular orbit with 
a radial-velocity semi-amplitude of 67\,m\,s$^{\rm -1}$. Their 
inferred minimum mass for the companion is 
$m_{\rm 2}\sin i$\,=\,2.26$\pm$\,0.18\,${\mathrm M_{\rm Jup}}$ 
(assuming $M_{\rm 1}$\,=\,1.03\,${\mathrm M_{\sun}}$).

\subsubsection{CORALIE data and orbital solution for $\iota$\,Hor}

We have 26 {\footnotesize CORALIE} radial-velocity measurements of 
$\iota$\,Hor. The time interval between the first and the last 
measurement is a little less than two orbital cycles 
($\Delta$\,t=\,604\,days) and we are able to compute our own orbital 
solution. The mean photon noise error of these measurements is 
6.5\,m\,s$^{-1}$. This rather high value for such a bright star 
($m_{\rm V}$\,=\,5.4) is due to the rather large rotational velocity 
of $\iota$\,Hor: $v\sin i$\,=\,6.1\,$\pm$\,1\,km\,s$^{-1}$ 
\citep[{\footnotesize CORAVEL} data, calibration by ][]{Benz84} in 
agreement with \citet{Saar97} who found 
$v\sin i$\,=\,5.7\,$\pm$\,0.5\,km\,s$^{-1}$. Table\,\ref{iotahororb} 
lists our fitted orbital elements compared to those published by 
\citet{Kurster2000}. We have fixed the orbital period to their value 
because of their much longer time coverage ($\simeq$\, 6 cycles). 
Although our solution is a little bit different, 
{\footnotesize CORALIE} measurements are compatible with the solution 
by \citet{Kurster2000}\footnote{Note that the $T_{\rm 0}$ published 
by K\"urster et al. corresponds to the time of maximum radial velocity 
whereas our $T$ is the time of closest approach. In our time system, 
the orbital phase of the maximum is $\phi$\,=\,0.745. We have:
$T$\,=\,$T_{\rm 0}$\,+\,(0.255\,+N$_{\rm cycles}$). In our case, 
N$_{\rm cycles}$\,=\,3. The $T$ value corresponding to their 
$T_{\rm 0}$ value is 51\,348.9 in agreement with our value}. The 
residuals from the fitted orbit obtained by imposing their period, 
eccentricity and semi-amplitude are 17\,m\,s$^{-1}$. Our orbital 
elements lead to a minimum planetary mass of 
$m_{\rm 2}\sin i$\,=\,2.41\,$\pm$\,0.18\,${\mathrm M_{\rm Jup}}$\ 
(assuming $M_{\rm 1}$\,=\,1.03\,${\mathrm M_{\sun}}$). The residuals 
from the fitted orbit are abnormally large 
($\sigma(O-C)$=16.6\,m\,s$^{-1}$, 
P($\chi^{\rm 2}$)\,=\,6$\cdot$10$^{\rm -6}$). Assuming a typical 
intrinsic precision of 7\,m\,s$^{-1}$\ for {\footnotesize CORALIE} and 
taking 6.5\,m\,s$^{-1}$\ for the statistical error, the expected 
residuals value is $\sigma(O-C)$\,$\simeq$\,10.3\,m\,s$^{-1}$ so the 
observed extra-scatter is of the order of 14\,m\,s$^{-1}$. These 
abnormally high residuals are also observed by \citet{Kurster2000}. 
Their observed residuals are 27\,m\,s$^{-1}$\ to be compared with 
their typical precision of 17\,m\,s$^{-1}$. $\iota$\,Hor\ is 
chromospherically active: $\log (R^{\prime}_{HK})$\,=\,$-$4.65 
\citep{HenryT96}. An excess scatter of 10\,$-$\,16\,m\,s$^{-1}$ is 
expected for this activity level and for this type of star 
\citep{Saar98,Santosact} so activity-related processes are probably 
responsible for the high residuals observed by both groups.

\begin{figure*}[th!]
    \psfig{width=0.85\hsize,file=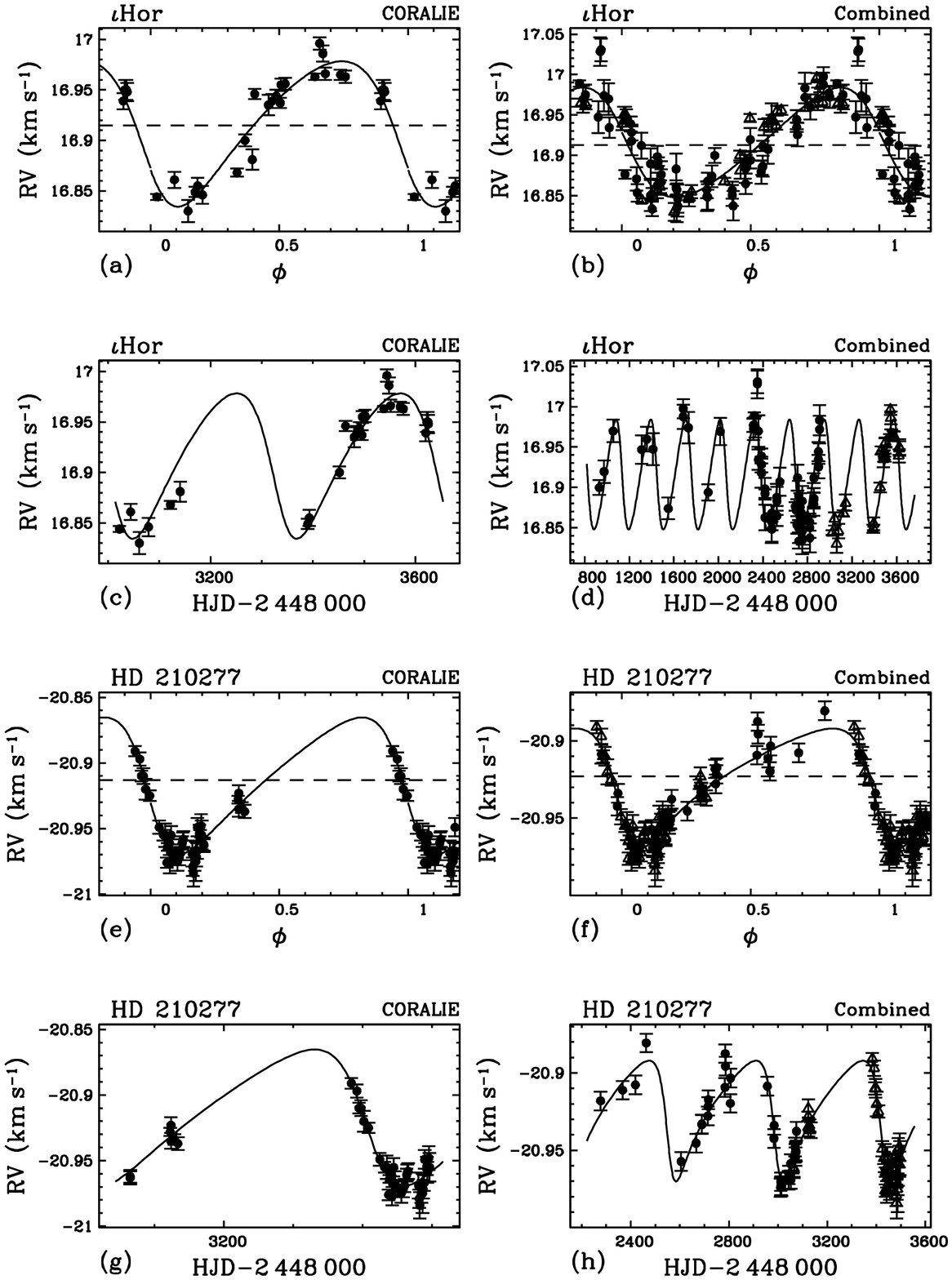}
    \caption{
    \label{iotahd210277orb}
    Phase-folded curves and temporal radial velocities for 
    $\iota$\,Hor and HD\,210277. Left column: CORALIE orbital solutions. 
    Right column: combined orbital solutions. {\bf (b)} and {\bf (d)}: 
    CORALIE (open triangles) and \citet[ filled dots]{Kurster2000}. 
    {\bf (f)} and {\bf (h)}: CORALIE (open triangles) and 
    \citet[ filled dots]{Vogt2000}}
\end{figure*}

Figure\,\ref{iotahd210277orb}a shows the {\footnotesize CORALIE} 
phase-folded curve with P fixed to 320.1\,days. The temporal 
velocities with the same orbital solution are displayed in 
Fig.\,\ref{iotahd210277orb}c.

\subsubsection{Combined orbital solution}

We present the orbital solution we obtain for $\iota$\,Hor\ after 
combining the {\footnotesize CORALIE} measurements with the velocities 
obtained by \citet{Kurster2000}. The mean uncertainty of their radial 
velocities is about 17.5\,m\,s$^{-1}$\ a much larger value than the 
6.5\,m\,s$^{-1}$\ of the {\footnotesize CORALIE} measurements. We used 
only the best K\"urster et al. velocities for fitting the combined 
orbit (data points with $\epsilon _{\rm RV}$\,$<$\,$20$\,m\,s$^{-1}$). 
62 measurements matched this criterion. K\"urster et al. measurements 
are differential velocities (velocity difference between the observed 
spectra and a reference spectra) so they have an arbitrary 
systemic velocity. Our computed combined orbital solution included 
the velocity offset between the two sets of data as an additional 
free parameter. The fitted offset is (C: {\footnotesize CORALIE}, K: 
K\"urster): $\Delta${\footnotesize RV} = {\footnotesize RV}$_{\rm C}$ 
$-$ {\footnotesize RV}$_{\rm K}$ = 16.910 $\pm$ 0.010 km\,s$^{-1}$.

Table\,\ref{iotahororb} provides the fitted combined orbital solution. 
Figure\,\ref{iotahd210277orb}b shows the combined phase-folded curve. 
The temporal velocities with the combined solution are displayed in 
Fig.\,\ref{iotahd210277orb}d. The computed planetary minimum mass in 
this case is 
$m_{\rm 2}\sin i$\,=\,2.24\,$\pm$\,0.13\,${\mathrm M_{\rm Jup}}$\footnote{The 
combined orbital solution has a more uncertain periastron epoch than 
the K\"urster et al. time of maximum radial-velocity. We fitted the 
orbital solution to the K\"urster et al. velocities using our 
software. We obtained a solution compatible with the K\"urster et al. 
published solution but with a 12\,days uncertainty on this parameter. 
The difference between the uncertainties is possibly due to 
differences in the fitting processes}.

\subsection{The planetary companion of HD\,210277}

The presence of a planetary companion orbiting the late G dwarf 
{\footnotesize HD}\,210277 has been announced a few years ago by 
\citet{Marcy99}. In a recent paper, \citet{Vogt2000} have presented an 
updated version of the orbital solution. To obtain this new orbit, 
these authors have added new measurements with the 
{\footnotesize HIRES} spectrograph \citep{Vogt1994} on the 
{\footnotesize KECK} telescope. Improvements in their reduction 
software have also increased their precision. Our 42 radial-velocity 
measurements are compatible with the orbital solution by 
\citet{Vogt2000}.

\begin{table*}[th!]
\caption{
\label{iotahororb} 
Orbital elements for $\iota$\,Hor. Results by \citet{Kurster2000} are 
shown for comparison. The orbital solution combining CORALIE (C) and 
K\"urster et al. (K) data is also presented. T$_{\rm max}$ is the 
time of maximum radial velocity}
\begin{tabular}{ll|r@{  $\,\pm\,$  }lr@{  $\,\pm\,$  }lr@{  $\,\pm\,$  }l}
\hline
\noalign{\vspace{0.025cm}}
{\bf Parameter}            & {\bf Units}                             & \multicolumn{2}{c}{\bf {\footnotesize CORALIE}} & \multicolumn{2}{c}{\bf K\"urster et al.} & \multicolumn{2}{c}{\bf Combined Solution}\\
\noalign{\vspace{0.025cm}}
\hline
\noalign{\vspace{0.025cm}}
$P$                        & (days)                                  & \multicolumn{2}{c}{320.1 (fixed)}               & 320.1         & 2.1                      & 311.3         & 1.3                      \\
$T$                        & ({\footnotesize HJD}\,$-$\,2\,400\,000) &    51\,334  & 14                                & \multicolumn{2}{c}{...}                  &     51\,308.8 & 10.4                     \\
$T_{\rm max}$              & ({\footnotesize HJD}\,$-$\,2\,400\,000) &  \multicolumn{2}{c}{...}                        & 50\,307.0     & 3.0	                  & \multicolumn{2}{c}{...}		     \\
$e$                        &                                         & 0.25        & 0.07                              & 0.161         & 0.069                    & 0.22          & 0.06                     \\
$\gamma$                   & (km\,s$^{-1}$)                          & 16.915      & 0.003                             & \multicolumn{2}{c}{...}                  & 16.913        & 0.003                    \\
$\omega$                   & ($\degr$)                               & 119         & 77                                & 83            & 11                       & 78.9          & 13.1                     \\
$K_{\rm 1}$                & (m\,s$^{-1}$)                           & 72          & 5                                 & 67.0          & 5.1                      & 68            & 4                        \\
$a_{\rm 1} \sin i$         & ($10^{-3}$AU)                           & 2.06        & 0.15                              & 1.94          & ...                      & 1.88          & 0.11                     \\
$f_{\rm 1}(m)$             & ($10^{-9}$${\mathrm M_{\odot}}$)        & 11.5        & 2.5                               & 9.5           & ...                      & 9.5           & 1.6                      \\
$m_{\rm 2}\sin i$          & (${\mathrm M_{\rm Jup}}$)               & 2.41        & 0.18                              & 2.26          & 0.18                     & 2.24          & 0.13                     \\
\noalign{\vspace{0.025cm}}
\hline
\noalign{\vspace{0.025cm}}
$N_{\rm meas}$             &                                         & \multicolumn{2}{c}{26}                          & \multicolumn{2}{c}{95}                   & \multicolumn{2}{c}{88 (C:26, K:62)}       \\ 
$\sigma(O-C)$              & (m\,s$^{-1}$)                           & \multicolumn{2}{c}{16.6}                        & \multicolumn{2}{c}{27.0}                 & \multicolumn{2}{c}{23.2 (C:20.2, K:24.3)} \\    
$\chi^{\rm 2}$             &                                         & \multicolumn{2}{c}{62.15}                       & \multicolumn{2}{c}{...}                  & \multicolumn{2}{c}{186.87}\\
P($\chi^{\rm 2}$)          &                                         & \multicolumn{2}{c}{6$\cdot$10$^{-6}$}           & \multicolumn{2}{c}{...}                  & \multicolumn{2}{c}{$<$\,5$\cdot$10$^{-7}$}\\    
\noalign{\vspace{0.025cm}}
\hline
\end{tabular}
\end{table*}

\begin{table*}[th!]
\caption{
\label{hd210277orb_par} 
Orbital elements for HD\,210277. Results by \citet{Vogt2000} are shown 
for comparison. The orbital solution combining CORALIE (C) and Vogt et 
al. (V) data is also presented}
\begin{tabular}{ll|r@{  $\,\pm\,$  }lr@{  $\,\pm\,$  }lr@{  $\,\pm\,$}l  }
\hline
\noalign{\vspace{0.025cm}}
{\bf Parameter}   & {\bf Units}                             & \multicolumn{2}{c}{\bf {\footnotesize CORALIE}} & \multicolumn{2}{c}{\bf Vogt et al.} & \multicolumn{2}{c}{\bf Combined Solution}\\
\noalign{\vspace{0.025cm}}
\hline
\noalign{\vspace{0.025cm}}
$P$               & (days)                                  & \multicolumn{2}{c}{436.6 (fixed)}               & 436.6   & 4                         & 435.6     & 1.9\\
$T$               & ({\footnotesize HJD}\,$-$\,2\,400\,000) &    51\,410  & 13                                & 51\,428 & 4                         & 51\,426.4 & 2.5\\
$e$               &                                         & 0.342       & 0.055                             & 0.45    & 0.03                      & 0.450     & 0.015\\
$\gamma$          & (km\,s$^{-1}$)                          & $-$20.913   & 0.006                             & ...     & ...                       & $-$20.923 & 0.001\\
$\omega$          & ($\degr$)                               & 104         & 13                                & 118     & 6                         & 117.4     & 3.3\\ 
$K_{\rm 1}$       & (m\,s$^{-1}$)                           & 52          & 9                                 & 39      & 2                         & 39        & 1\\
$a_{\rm 1} \sin i$& ($10^{-3}$AU)                           & 1.96        & 0.33                              & 1.39    & ...                       & 1.40      & 0.03\\
$f_{\rm 1}(m)$    & ($10^{-9}$M$_{\odot}$)                  & 5.3         & 2.7                               & 1.91    & ...                       & 1.97      & 0.13\\
$m_{\rm 2}\sin i$ & (M$_{\rm Jup}$)                         & 1.73        & 0.29                              & 1.23    & ...                       & 1.24      & 0.03\\
\noalign{\vspace{0.025cm}}
\hline
\noalign{\vspace{0.025cm}}
$N_{\rm meas}$    &                                         & \multicolumn{2}{c}{42}                          & \multicolumn{2}{c}{45}              & \multicolumn{2}{c}{87 (C:42, V:45)}\\ 
$\sigma(O-C)$     & (m\,s$^{-1}$)                           & \multicolumn{2}{c}{7.5}                         & \multicolumn{2}{c}{3.24}            & \multicolumn{2}{c}{6.10 (C:8.23, V:3.23)}\\
$\chi^{\rm 2}$    &                                         & \multicolumn{2}{c}{24.45}                       & \multicolumn{2}{c}{...}             & \multicolumn{2}{c}{61.59}\\
P($\chi^{\rm 2}$) &                                         & \multicolumn{2}{c}{0.944}                       & \multicolumn{2}{c}{...}             & \multicolumn{2}{c}{0.937}\\
\noalign{\vspace{0.025cm}}
\hline
\end{tabular}
\end{table*}

\subsubsection{CORALIE data for HD\,210277}

42 {\footnotesize CORALIE} radial velocities of 
{\footnotesize HD}\,210277 were measured between 
{\footnotesize HJD}\,=2\,451\,064 and 
{\footnotesize HJD}\,=2\,451\,497. The mean photon noise error of 
these measurements is 6\,m\,s$^{-1}$. With 433 days between the first 
and the last measurement, our time span is a little shorter than the 
orbital period published by Vogt et al. (436.6\,days) so we do not 
have a reliable constraint on the periodicity of the system. We choose 
to set $P$\,=\,436.6\,days as a fixed parameter in our fit. 
Table\,\ref{hd210277orb_par} shows our fitted orbital solution 
compared to the one by \citet{Vogt2000}. There are some small 
differences between the two solutions (larger semi-amplitude and 
smaller eccentricity for the {\footnotesize CORALIE} solution) but 
they are mostly due to our poor phase coverage (no data point at the 
phase of the maximum radial velocity). If we set constant in the fit 
all the orbital elements by Vogt et al. (except for the 
$\gamma$\,velocity which we keep as a free parameter), we obtain a 
very good agreement between our data and their solution. We find 
$\sigma(O-C)$\,=\,7.7\,m\,s$^{-1}$\ to be compared with the 
7.5\,m\,s$^{-1}$\ obtained fixing only the period. Assuming 
0.92\,${\mathrm M_{\sun}}$ for the mass of the primary 
\citep{Gonzalez99}, we find a planetary minimum mass of 
$m_{\rm 2}\sin i$\,=\,1.73\,$\pm$\,0.29\,${\mathrm M_{\rm Jup}}$, a 
value significantly higher than the one found by Vogt et al. 
(1.23\,${\mathrm M_{\rm Jup}}$). This difference is due to the 
difference in the velocity semi-amplitude (52\,m\,s$^{-1}$\ for 
{\footnotesize CORALIE}, 39\,m\,s$^{-1}$\ for {\footnotesize HIRES}).

Figure\,\ref{iotahd210277orb}e shows the phase-folded curve obtained 
in this case for {\footnotesize HD}\,210277. The temporal velocities 
with the same orbital solution are displayed in 
Fig.\,\ref{iotahd210277orb}g.

\begin{table*}[th!]
\caption{
\label{hd217107sol}
Orbital elements for HD\,217107. Results by \citet{Vogt2000} are shown 
for comparison. The systemic drift was taken into account in both 
cases. The orbital solution combining CORALIE data (C) with the 
measurements by \cite{Fischer99} (F) and \citet{Vogt2000} (V) is also 
presented. There are no significant differences between these 
solutions}
\begin{tabular}{ll|r@{  $\,\pm\,$  }lr@{  $\,\pm\,$  }lr@{  $\,\pm\,$  }l}
\hline
\noalign{\vspace{0.025cm}}
{\bf Parameter}              & {\bf Units}                             & \multicolumn{2}{c}{\bf {\footnotesize CORALIE}}  & \multicolumn{2}{c}{\bf Vogt et al.}   & \multicolumn{2}{c}{\bf Combined solution}\\
                             &                                         & \multicolumn{2}{c}{Time bins}                    & \multicolumn{2}{c}{Linear correction} & \multicolumn{2}{c}{Time bins}\\
\noalign{\vspace{0.025cm}}
\hline
\noalign{\vspace{0.025cm}}
$P$                          & (days)                                  & 7.1260                 & 0.0005                  & 7.1260    & 0.0007                    & 7.1262                 & 0.0003\\
$T$                          & ({\footnotesize HJD}\,$-$\,2\,400\,000) & 51\,452.388            & 0.079                   & 51\,331.4 & 0.2                       & 51\,452.477            & 0.056\\
$e$                          &                                         & 0.126                  & 0.009                   & 0.140     & 0.02                      & 0.134$^{\dagger}$     & 0.007\\
$\gamma$                     & (km\,s$^{-1}$)                          & $-$13.413$^{\natural}$ & 0.001                   & ...       & ...                       & $-$13.413$^{\natural}$ & 0.001\\
$\omega$                     & ($\degr$)                               & 24                     & 4                       & 31        & 7                         & 29                     & 3\\ 
$K_{\rm 1}$                  & (m\,s$^{-1}$)                           & 140                    & 1                       & 140       & 3                         & 140                    & 1\\
$a_{\rm 1} \sin i$           & ($10^{-5}$AU)                           & 9.1                    & 0.1                     & 9.1       & ...                       & 9.07                   & 0.07\\
$f_{\rm 1}(m)$               & ($10^{-9}$M$_{\odot}$)                  & 1.96                   & 0.06                    & 1.97      & ...                       & 1.99                   & 0.04\\
$m_{\rm 2}\sin i$	     & (M$_{\rm Jup}$)                         & 1.275                  & 0.013                   & 1.28      & 0.4                       & 1.282                  & 0.009\\
$N_{\rm meas}$               &                                         & \multicolumn{2}{c}{63}                           & \multicolumn{2}{c}{21}                & \multicolumn{2}{c}{98 (C:63, F:14, V:21)}\\ 
$\sigma(O-C)$                & (m\,s$^{-1}$)                           & \multicolumn{2}{c}{7.0}                          & \multicolumn{2}{c}{4.2}               & \multicolumn{2}{c}{7.21 (C:7.40, F:10.18, V:3.73)}\\    
\noalign{\vspace{0.025cm}}
\hline
\noalign{\vspace{0.025cm}}
\end{tabular}
\\
$^{\dagger}$ highly significant according to the Lucy \& Sweeney test \citep{Lucy}\\
$^{\natural}$ at {\footnotesize HJD}\,=\,2\,451\,450 (= median of the third {\footnotesize CORALIE} time bin)
\end{table*}

\subsubsection{Combined orbital solution}

In order to obtain better constraints on the orbital elements, we have 
combined the {\footnotesize CORALIE} data with the 
{\footnotesize HIRES} radial-velocity measurements by \citet{Vogt2000}. 
As for the \citet{Kurster2000} velocities, the Vogt et al. 
measurements are differential velocities so they have an arbitrary 
systemic velocity. We first fitted their systemic velocity (Keplerian 
fit with $\gamma$ as the only free parameter, other parameters fixed 
to the Vogt et al. values). We made the same computation with the 
{\footnotesize CORALIE} velocities in order to compute the velocity 
offset between the two sets of data. We found (C: 
{\footnotesize CORALIE}, V: Vogt et al.): 
{\footnotesize RV}$_{\rm C}$\,=\,{\footnotesize RV}$_{V}$\,$-$\,20.930 km\,s$^{-1}$

The combined orbital solution, whose results are presented in 
Table\,\ref{hd210277orb_par}, was obtained by fitting a Keplerian 
orbit to the combined data ({\footnotesize CORALIE}\,+\,corrected 
{\footnotesize HIRES} velocities). We checked our velocity correction 
by fitting the velocity offset (additional free parameter). The fitted 
residual offset is smaller than 0.5\,m\,s$^{-1}$. The computed 
minimum mass is 
$m_{\rm 2}\sin i$\,=1.24\,$\pm$\,0.03\,${\mathrm M_{\rm Jup}}$. 
Figure\,\ref{iotahd210277orb} shows the phase-folded-curve (panel f) 
and the velocities as a function of time (panel h) for the 
{\footnotesize CORALIE} + {\footnotesize HIRES} combined solution.

\begin{figure*}[th!]
    \psfig{width=0.85\hsize,file=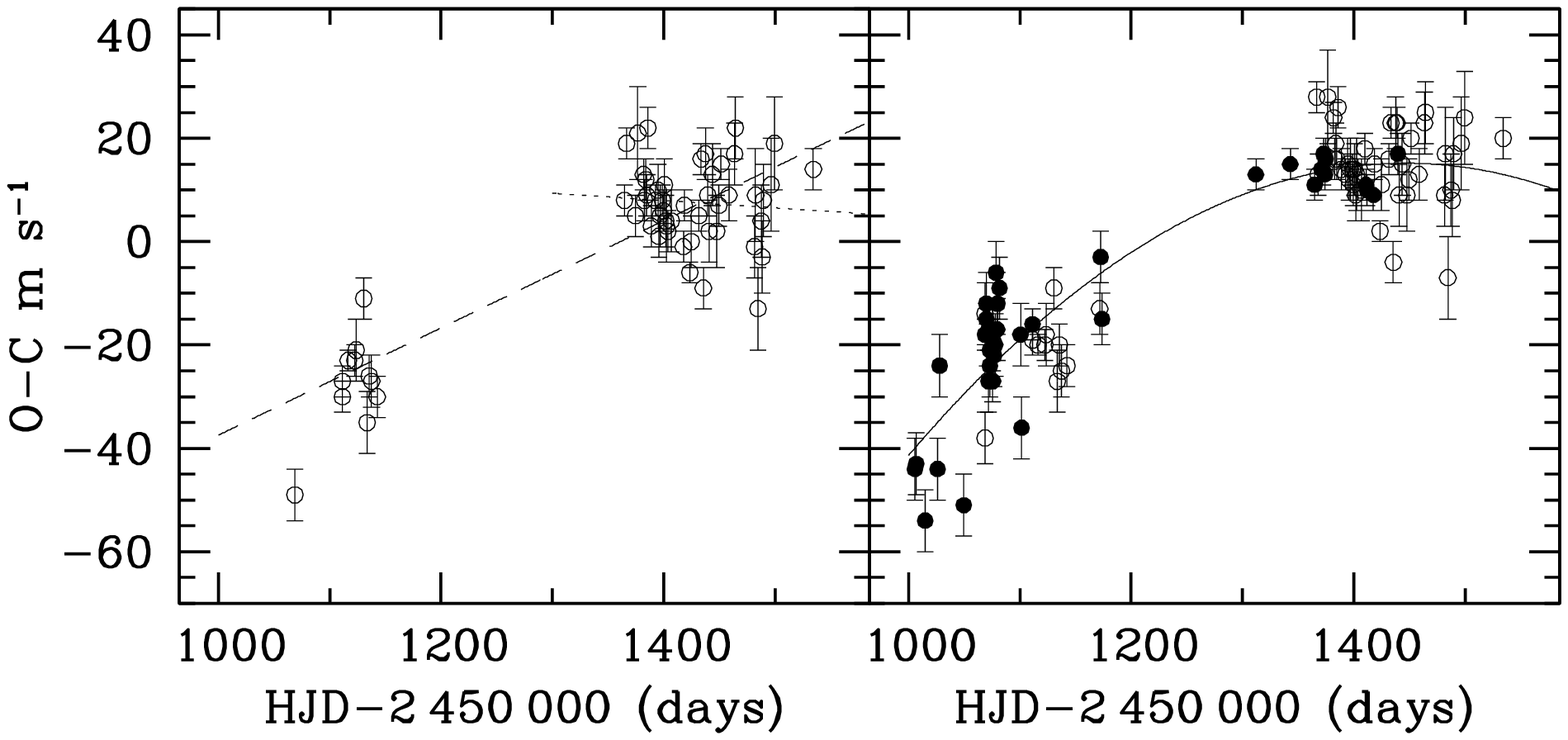}
    \caption{
    \label{hd217107res}
    Left: CORALIE data. Residuals versus time. The orbital solution by 
    \citet{Vogt2000} has been used. Only our 58 best measurements are 
    plotted. A linear fit to these residuals gives a slope of 
    37.9\,$\pm$\,3.4\,\,m\,s$^{-1}$\,yr$^{-1}$ (dashed line). A linear 
    fit to the most recent data (data from the last observing season 
    only) gives a slope of 
    -5\,$\pm$\,10\,\,m\,s$^{-1}$\,yr$^{-1}$ (dotted line). This 
    difference between the two slopes tend to indicate a non-linear 
    trend. 
    Right: residual from the combined orbit 
    (see Table\,\ref{hd217107sol}) of HD\,217107. Open dots: CORALIE 
    data, Filled dots: \citet{Vogt2000} and \citet{Fischer99} data. 
    The curve shows the result of a quadratic fit to these residuals. 
    The fitted function is: 
    O$-$C($\tau$)\,=\,$K_{\rm 0}$\,$+$\,$K_{\rm 1}$$\tau$\,$+$\,$K_{\rm 2}$$\tau^2$ 
    where $\tau$\,=\,HJD\,$-$\,2\,450\,000. The quadratic term is 
    significant:
    $K_{\rm 2}$\,=\,$-$2.83\,$\pm$\,0.54 (10$^{-4}$\,m\,s$^{-1}$\,d$^{-2}$). 
    The $\chi^{2}$ statistics shows that a linear trend is very 
    unlikely (see Table\,\ref{xi2})}
\end{figure*}

\subsection{The planetary companion of HD\,217107}\label{217107}

The discovery of a planetary companion orbiting the G7 dwarf 
{\footnotesize HD}\,217107 was initially announced by 
\citet{Fischer99}. This discovery was based on radial velocities 
obtained with the {\footnotesize HAMILTON} echelle spectrometer 
\citep{Vogt1987} at Lick Observatory and with the 
{\footnotesize HIRES} echelle spectrometer \citep{Vogt1994} on the 
{\footnotesize KECK} telescope. In a recent paper, \citet{Vogt2000} 
have presented updated orbital elements for this planet. This orbit 
was based on radial-velocity measured with the {\footnotesize HIRES} 
spectrometer \citep{Vogt1994} and agreed with the orbital solution 
from the discovery paper. \citet{Vogt2000} also announced the 
discovery of a linear drift of the systemic velocity with a slope of 
$39.4$\,$\pm$\,$3.5$\,\,m\,s$^{-1}$\,yr$^{-1}$. This trend was later 
confirmed by Fischer et al. using their most recent observations 
(slope\,=\,$43.3$\,$\pm$\,2.8\,\,m\,s$^{-1}$\,yr$^{-1}$). 
In Sect.\,\ref{hd217107Cdos}, we present the {\footnotesize CORALIE} 
radial-velocity measurements and orbital solution for 
{\footnotesize HD}\,217107\ confirming the orbital solutions by 
\citet{Fischer99} and \citet{Vogt2000}. We also confirm the 
$\gamma$\,velocity drift revealing the presence of a third massive 
body in the system. We show that this drift is probably not linear. 
An orbital solution combining all the available data is presented in 
Sect.\,\ref{hd217107combsec}. 

\subsubsection{CORALIE data for HD\,217107}\label{hd217107Cdos}

In this section we present the {\footnotesize CORALIE} data for 
{\footnotesize HD}\,217107. We shall focus on the systemic velocity 
drift correction in order to adjust the best orbital solution. 

63 {\footnotesize CORALIE} radial-velocity measurements have been 
gathered during the first two observing seasons (1998--1999). These 
velocities span from {\footnotesize HJD}\,=\,2\,451\,538 to 
{\footnotesize HJD}\,=\,2\,451\,067) and have a mean photonic error of 
5.5\,m\,s$^{-1}$.

We first imposed the solution by \citet{Vogt2000} to our 58 best 
measurements to see the behaviour of the residuals (we only fitted the 
$\gamma$\,velocity). The residuals obtained from this solution are 
16.1\,m\,s$^{-1}$, a clearly abnormally high value. 

Figure\,\ref{hd217107res} (left) shows the residuals as a function of 
time. A trend is clearly visible. A linear fit to the residuals of 
the two seasons gives a slope of 
37.9\,$\pm$\,3.4\,\,m\,s$^{-1}$\,yr$^{-1}$ (compatible with the 
previously detected trends). A linear fit to the residuals of the 
last observing season ({\footnotesize HJD}\,$>$\,2\,451\,300) gives a 
slope of $-$5\,$\pm$\,10\,\,m\,s$^{-1}$\,yr$^{-1}$ (consistent with 
zero). These two slope values are not compatible. The slope 
difference is not highly significant (P($\chi^{\rm 2}$)\,=\,0.29 and 
0.85, respectively for the last season measurements). At this point 
and using the {\footnotesize CORALIE} data alone, there are only 
clues indicating that the drift is not linear over the two seasons.

\begin{figure*}[th!]
    \psfig{width=0.85\hsize,file=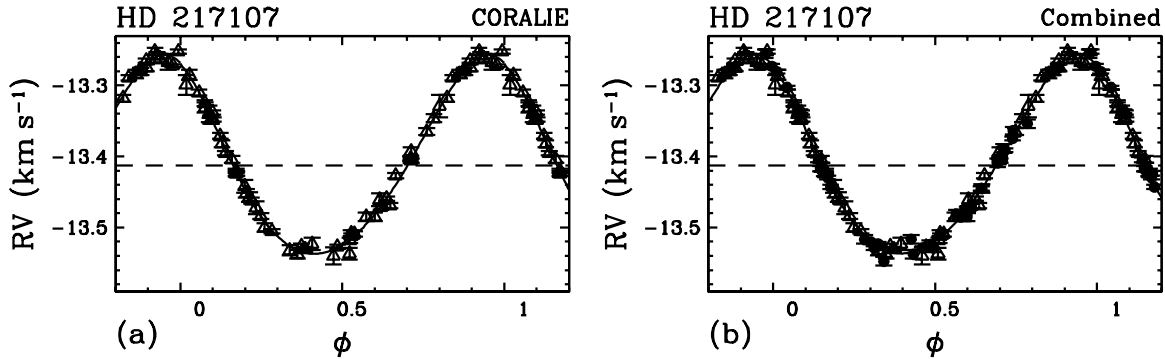}
    \caption{
    \label{hd217107orbplots}
    Left: CORALIE phase-folded curve. 
    Right: CORALIE, \citet{Vogt2000} and \citet{Fischer99} 
    combined phase-folded curve. A time bins correction has been used 
    in both cases. A linear fit to the residuals around these two 
    orbital solutions gives a flat slope}
\end{figure*}

The {\footnotesize CORALIE} orbital solution presented in 
Table\,\ref{hd217107sol} was obtained after correcting the 
$\gamma$\,velocity drift using a time bins correction. We split our 
data into three time bins (see Table\,\ref{bins1}). We fit a Keplerian 
orbit to the data in the third time bin. We fit the systemic velocity 
of the second time bin (Keplerian fit with $\gamma$ as the only free 
parameter, other parameters fixed to the fitted values for the third 
time bin). We correct the velocities of the second time bin so that 
they have the same systemic velocity as the velocities of the third 
time bin. We then fit a new Keplerian orbit to the velocities of the 
third time bin plus the corrected velocities of the second time bin. 
We correct the velocities of the first time bin using the same procedure. 
Once all the bins are corrected, we fit a Keplerian orbit to the full set 
of data. With this kind of drift correction, we do not have to make 
any assumption on the drift shape. The residuals from the fitted 
solution (7\,m\,s$^{-1}$) are compatible with the typical precision 
of {\footnotesize CORALIE}. We also computed an orbital solution using 
a linear correction. The residual from this solution are substantially 
higher: 9.1\,m\,s$^{-1}$.

In Table\,\ref{hd217107sol} are listed the fitted parameters for the 
time bins correction. The orbital solution by  \citet{Vogt2000}, is 
shown for comparison. There are no substantial differences between 
these solutions. Assuming a primary mass 
$M_{\rm 1}$\,=\,0.96\,${\mathrm M_{\sun}}$ (similarly to 
\citealt{Fischer99}), the computed minimum mass is 
$m_{\rm 2}\sin i$\,=\,1.275\,$\pm$\,0.013\,${\mathrm M_{\rm Jup}}$.
Figure\,\ref{hd217107orbplots}a shows the {\footnotesize CORALIE} 
phase-folded curve for the time bins correction.

\begin{table}[t!]
\caption[]{\label{bins1}Time bins correction for the CORALIE 
measurements and for the data by \citet{Fischer99} and 
\citet{Vogt2000}. All the velocities are in km\,s$^{-1}$}
\begin{tabular}{cccr@{  $\,\pm\,$  }lc}
\hline
\noalign{\vspace{0.025cm}}
{\bf {\footnotesize JD}$_{\rm min}^{\dagger}$} & {\bf {\footnotesize JD}$_{\rm max}^{\dagger}$} & {\bf N$_{\rm meas}$} & \multicolumn{2}{c}{{\bf $\gamma$}} & {\bf $\Delta$RV$^{\ddagger}$}\\
\noalign{\vspace{0.025cm}}
\hline
\hline
\noalign{\vspace{0.025cm}}
\multicolumn{6}{c}{{\footnotesize CORALIE}}\\
\noalign{\vspace{0.025cm}}
\hline
\noalign{\vspace{0.025cm}}
67                                             & 69                                             & 2                    & $-$13.473 & 0.012                  & 0.060\\
111                                            & 143                                            & 10                   & $-$13.448 & 0.002                  & 0.035\\
364                                            & 538                                            & 51                   & $-$13.413 & 0.001                  & 0\\
\noalign{\vspace{0.025cm}}
\hline
\noalign{\vspace{0.025cm}}
\multicolumn{6}{c}{\citet{Fischer99}}\\
\noalign{\vspace{0.025cm}}
\noalign{\vspace{0.025cm}}
\hline
5                                              & 50                                             & 6                    &   0.042   & 0.004                  & $-$13.455\\
76                                             & 102                                            & 8                    &   0.068   & 0.004                  & $-$13.481\\
\noalign{\vspace{0.025cm}}
\hline
\noalign{\vspace{0.025cm}}
\multicolumn{6}{c}{\citet{Vogt2000}}\\
\noalign{\vspace{0.025cm}}
\hline
\noalign{\vspace{0.025cm}}
68                                             & 76                                             & 7                    &  $-$0.004 & 0.002                  & $-$13.409\\
171                                            & 174                                            & 3                    &     0.004 & 0.003                  & $-$13.417\\
312                                            & 375                                            & 7                    &     0.029 & 0.001                  & $-$13.442\\
410                                            & 439                                            & 4                    &     0.030 & 0.002                  & $-$13.443\\
\noalign{\vspace{0.025cm}}
\hline
\noalign{\vspace{0.025cm}}
\end{tabular}
\\
$^{\dagger}$ {\footnotesize JD\,=\,HJD}\,$-$\,2\,451\,000.
\\
$^{\ddagger}$ Correction to set the velocities into the systemic 
velocity of the third CORALIE  time bin
\end{table}
\begin{table}[t!]
\caption{\label{xi2} 
$\chi^{\rm 2}$ statistics for the different types of 
$\gamma$\,velocity correction for HD\,217107. $N_{\rm d.of.f}$ is 
the number of degrees of freedom}
\begin{tabular}{lcccc}
\hline
\noalign{\vspace{0.025cm}}
{\bf Correction} & {\bf $N_{\rm param}$} & {\bf $N_{\rm d.of.f}$} & {\bf $\chi^{\rm 2}$} & {\bf P($\chi^{\rm 2}$)} \\
\noalign{\vspace{0.025cm}}
\hline
\noalign{\vspace{0.025cm}}
Time bins        & 14                    & 84                     & 68.37                & 0.892\\
Quadratic        & 10                    & 88                     & 107.91               & 0.074\\
Linear           & 9                     & 89                     & 405.73               & $<$\,5$\cdot$10$^{\rm -7}$\\
\noalign{\vspace{0.025cm}}
\hline
\end{tabular}
\end{table}

\subsubsection{Combined orbital solution}\label{hd217107combsec}

In order to improve the precision on the orbital elements, we computed 
a combined orbital solution using the {\footnotesize CORALIE} data, 
the velocities by \citet{Fischer99} and the velocities by 
\citet{Vogt2000}. Before computing the combined orbital solution, we 
had to bring the different sets of data into a common velocity system. 
We decided to set all the velocities into the {\footnotesize CORALIE} 
$\gamma$\,velocity at {\footnotesize HJD}\,=\,2\,451\,450 reference 
frame (median of the third time bin, see Sect.\,\ref{hd217107Cdos}). 
We applied a time bins correction for the data by \citet{Fischer99} 
and by \citet{Vogt2000} in order to remove simultaneously the systemic 
drift effect and the radial-velocity offsets (differential velocities) 
between the different instruments.
 
The fitted orbital elements for the combined solution are listed in 
Table\,\ref{hd217107sol}. The computed minimum mass of the planet is 
$m_{\rm 2}\sin i$\,=\,1.282\,$\pm$\,0.009\,${\mathrm M_{\rm Jup}}$. 
Constraining the eccentricity to be zero causes the residuals to 
increase to 14.4\,m\,s$^{-1}$ indicating that the non-zero 
eccentricity is highly significant. The confidence level as computed 
with the Lucy \& Sweeney test \citep{Lucy} is almost 100\,\%. 
Figure\,\ref{hd217107orbplots}b shows the phase-folded curve for the 
combined orbital solution.

Using the uncorrected data, we now want to describe more precisely the 
shape of the systemic drift. We first have to set all the available 
data into the same velocity system. A time bins correction can not be 
used in this case because it cancels out all the $\gamma$\,velocity 
information (instrumental offsets and drift effect). The offset 
between the velocities by \citet{Fischer99} and \citet{Vogt2000} is 
corrected using 7 measurements present in the two data sets. The 
offset between the {\footnotesize CORALIE} and the \citet{Vogt2000} 
velocities is obtained by computing a combined orbital solution using 
data taken during a common epoch (namely, the beginning of the second 
season) with the offset as an additional free parameter. We have (C: 
{\footnotesize CORALIE}, F: Fischer et al., V: Vogt et al.):
\begin{eqnarray*}
RV_{\rm V}  & = & RV_{\rm F} - 70.5   \pm 2.3\,\mathrm{m\,s}^{-1}\\
RV_{\rm C}  & = & RV_{\rm V} - 13.442 \pm 0.010\,\mathrm{km\,s}^{-1}
\end{eqnarray*}
All the velocities have been finally set into the 
{\footnotesize CORALIE} system. Figure\,\ref{hd217107res} (right) 
shows the residuals from the combined orbit (the combined orbit of 
Table\,\ref{hd217107sol} has been used with the $\gamma$\,velocity as 
the only free parameter). The velocity drift is clearly not linear. A 
quadratic fit to these residuals has been computed and is also plotted 
in Fig.\,\ref{hd217107res} (right, solid line). The second order term 
of this fit is significant: 
$-$2.83\,$\pm$\,0.54\,(10$^{-4}$\,m\,s$^{-1}$\,d$^{-2}$).
The available data do not give a reliable constraint on the mass of 
the third body of the system. For the moment, a companion within the 
planetary mass range cannot be ruled out.

Finally, if we fit a Keplerian orbit to the velocities corrected for 
the quadratic drift, we end up with a solution compatible with the 
combined solution of Table\,\ref{hd217107sol}. The residuals from this 
solution are 8.86\,m\,s$^{-1}$. Table\,\ref{xi2} shows the values of 
the $\chi^{\rm 2}$ statistics for the different corrections of the 
$\gamma$\,velocity drift. The huge difference in P($\chi^{\rm 2}$) 
between the linear and the quadratic corrections shows that a linear 
drift is very unlikely.

\section{Conclusion}

In this paper, we presented our {\footnotesize CORALIE} 
radial-velocity measurements for the 6 stars {\footnotesize GJ}\,3021, 
{\footnotesize HD}\,52265, {\footnotesize HD}\,169830, $\iota$\,Hor, 
{\footnotesize HD}\,210277 and {\footnotesize HD}\,217107. We reported 
the detection of 3 new planetary companions orbiting the solar type 
stars {\footnotesize GJ}\,3021, {\footnotesize HD}\,52265 and 
{\footnotesize HD}\,169830. These planets have longer periods than 
the "Hot Jupiters" (the periods are respectively 133.7, 119.2 and 
229.9 days). All the fitted orbits are elongated (the eccentricities 
are respectively 0.51, 0.35 and 0.35). They are all Jovian planets 
(the minimum masses are respectively 3.37, 1.05 and 
2.94\,${\mathrm M_{\rm Jup}}$). Using {\footnotesize CORALIE} data, 
we have estimated the chromospheric activity level of these candidates. 
{\footnotesize GJ}\,3021 is a very active star whereas 
{\footnotesize HD}\,52265 and {\footnotesize HD}\,169830 are quiescent. 
The high activity level of {\footnotesize GJ}\,3021 is probably 
responsible for the high residuals from the fitted orbit. We also 
estimated the Lithium abundances of these stars using 
{\footnotesize CORALIE} spectra. {\footnotesize GJ}\,3021 and 
{\footnotesize HD}\,52265 exhibit a strong 
\mbox{$\lambda$ 6707.8 \AA\ \ion{Li}{i}} absorption feature whereas 
no Lithium is detected for {\footnotesize HD}\,169830. The 3 stars 
have a higher metal content than the Sun. The spatial velocities of 
{\footnotesize GJ}\,3021, its metallicity, its chromospheric activity 
level and its Lithium abundance are consistent with the values 
observed for members of the Hyades Super-Cluster.

We confirmed the orbital solution by \citet{Kurster2000} for 
$\iota$\,Hor and the solutions by \citet{Vogt2000} for 
{\footnotesize HD}\,210277 and {\footnotesize HD}\,217107. For the 
latter, we also confirmed the systemic velocity drift showing it is 
probably not linear over our first two observing seasons. Finally, we 
computed combined orbital solutions for these 3 stars using all the 
available velocity measurements improving the precision on the orbital 
elements.

\begin{acknowledgements}
We acknowledge support from the Swiss National Research Foundation 
({\footnotesize FNRS}) and from the Geneva university. Support from 
Funda\c{c}\~{a}o para a Ci\^{e}ncia e Tecnologia, Portugal, to N.C.S. 
in the form of a scholarship is gratefully acknowledged. We wish to 
thank Fabien Carrier, Francesco Kienzle, Claudio Melo and Frederic 
Pont for additional measurements obtained during their own observing 
runs. This research has made use of the {\footnotesize SIMBAD} 
database, operated at {\footnotesize CDS}, Strasbourg, France.
\end{acknowledgements}

\bibliographystyle{apj}
\bibliography{MS10239}
\end{document}